\documentclass[a4paper,number,sort&compress,1p,preprint]{elsarticle}
\usepackage{graphicx}
\begin{document}
%%%%%%%%%%%%%%%%
%
%%%%%%%%%%%%%%%%%%%%%%%%%%%%%%%%%%%%%%%%%%%%%%%%%%%%%%%%%%%%%%%%%%%%%%
%%%%%%%%%%%%%%%%%%%%%%%%%%%%%%%%%%%%%%%%%%%%%%%%%%%%%%%%%%%%%%%%%%%%%%
%

\title{Lifetime of a target in the presence \\
       of {N} independent walkers}

\author{F\'elix Rojo\fnref{fn1}}
\ead{rojo@famaf.unc.edu.ar}

\author{Pedro A.\ Pury}
\ead{pury@famaf.unc.edu.ar}

\author{Carlos E.\ Budde\corref{cor1}}
\ead{budde@famaf.unc.edu.ar}

\fntext[fn1]{Fellow of Consejo Nacional de Investigaciones
Cient{\'{\i}}ficas y T{\'e}c\-nicas (CONICET)}
\cortext[cor1]{Corresponding author}

\address{
Facultad de Matem\'atica, Astronom\'\i a y F\'\i sica, \\
Universidad Nacional de C\'ordoba, \\
Ciudad Universitaria, X5000HUA C\'ordoba, Argentina}

%
%%%%%%%%%%%%%%%%%%%%%%%%%%%%%%%%%%%%%%%%%%%%%%%%%%%%%%%%%%%%%%%%%%%%%%
%
\begin{abstract}
We study the survival probability of an immobile target in presence
of N independent diffusing walkers. We address the problem of the
Mean Target Lifetime  and its dependence on the number and initial
distribution of the walkers when the trapping is perfect or imperfect.
We consider the diffusion on lattices and in the continuous space
and we address the bulk limit corresponding to a density of diffusing
particles and only one isolated trap.
Also, we use intermittent motion for optimization of search
strategies.
\end{abstract}
%
%%%%%%%%%%%%%%%%%%%%%%%%%%%%%%%%%%%%%%%%%%%%%%%%%%%%%%%%%%%%%%%%%%%%%%
%
%
%%%%%%%%%%%%%%%%%%%%%%%%%%%%%%%%%%%%%%%%%%%%%%%%%%%%%%%%%%%%%%%%%%%%%%
\maketitle
%%%%%%%%%%%%%%%%%%%%%%%%%%%%%%%%%%%%%%%%%%%%%%%%%%%%%%%%%%%%%%%%%%%%%%

%%%%%%%%%%%%%%%%%%%%%%%%%%%%%%%%%%%%%%%%%%%%%%%%%%%%%%%%%%%%%%%%%%%%%%
\section{Introduction}
\label{sec:intro}
%%%%%%%%%%%%%%%%%%%%%%%%%%%%%%%%%%%%%%%%%%%%%%%%%%%%%%%%%%%%%%%%%%%%%%

Several models of trapping of random walkers by a target have been
extensively discussed in the literature~\cite{010,020,030,040}.
In most cases, in the analysis of the target lifetime,
the initial distribution of walkers is taken at random
(spatially homogeneous distribution) and the trapping
is assumed ``perfect''~\cite{050,060,070}.
On the other hand, some results concerning to the {\em order
statistics} of a set of random walkers are known when they
are initially placed onto a given site of an Euclidean
lattice~\cite{080,090} or a fractal structure~\cite{020}.
Thus, the question that arises naturally is how different initial
configurations of the walkers, number of walkers, and trapping
mechanisms affect on the survival probability of the target.

In this paper we address the problem of trapping by a fixed
target at the origin in the presence of a set of $N$
independent random walkers.
The quantity of our interest is the lifetime of the trap, which
ends when {\em any} walker reaches the target under the appropriate
circumstances.
Our approach not only provides a unified framework that comprises
several situations of trapping scattered in the literature,
but also enables us to compute (analytical and numerically)
the lifetime of the trap in an exact way for a vast number
of practical situations.
Thus, we consider that the wandering of the walkers in the space
may be normal or anomalous diffusion, and that the trapping may
have particular characteristics:
It may be perfect, in which case the lifetime of the trap
reckons the time spend by the first walker to reach the origin,
or imperfect~\cite{060,100}. The last situation includes the
cases in which a walker passing by the origin is not trapped
with certainty.
Also, the present formalism could be applied even though we
have dynamical trapping, i.e., the state of trap change in time
like gated trapping~\cite{110,120}.

A novel application of our concepts comes from the hand of
intermittent motion phenomena, which has recently motivated
numerous studies in physics, chemistry,
and biology~\cite{122,124,126,127,128}.
In this piece of work, we show that the analysis of the trap's
lifetime, as a function of the transition rates among internal
states of the walker, allow us to optimize the intermittent
search strategy for a hidden target.

The paper is organized as follows. Section~\ref{sec:life} presents
the general formalism and define the concepts of survival of the
target, lifetime density, and mean quantities; and establishes
the connection with the problem of only one walker.
Particularly, the last issue or first-passage time problem~\cite{130},
is reviewed in Sec.~\ref{sec:FPT}, whereas in Sec.~\ref{sec:d-dim}
we reconsider the effects of dimensionality and number of walkers
on our problem.
This section also reviews the basic about the continuous--time
random walk (CTRW).
In Sec.~\ref{sec:initial} the effects due to initial spatial
distribution of the set walkers are analyzed and the bulk limit
is constructed.
Section~\ref{sec:illustrations} presents several assorted
illustrations for discrete and continuous systems with different
types of initial distributions and considering the effects
of finite size of the space and imperfection in the trapping
mechanism.
Last, in Sec.~\ref{sec:Intermittent}, we discuss the usefulness
of our approach for searching targets with stochastic intermittent
motion. Finally in Sec.\ref{sec:fin} we give our conclusions.

%%%%%%%%%%%%%%%%%%%%%%%%%%%%%%%%%%%%%%%%%%%%%%%%%%%%%%%%%%%%%%%%%%%%%%
\section{Lifetime of the Target}
\label{sec:life}
%%%%%%%%%%%%%%%%%%%%%%%%%%%%%%%%%%%%%%%%%%%%%%%%%%%%%%%%%%%%%%%%%%%%%%

The major objective of this contribution is the study of
the effects of the initial distribution of independent walkers
and the characteristics of the trapping process in the survival
of the target.
For this task, we begin reviewing and generalizing the formalism
developed in Ref.~\cite{050,060}.
The survival probability at time $t$, $\Phi_{N}(t)$, of the
static target (trap) at the origin in presence of $N$ independent
walkers that diffuse on a lattice can be written as~\cite{050}
\begin{equation}
\Phi_{N}(t) = \sum_{\vec{s}_1}\ldots\sum_{\vec{s}_N}
u(\vec{s}_1,..,\vec{s}_N) \prod_{i=1}^{N}\Phi_{1}(\vec{s}_i,t) \,,
\label{PhiN(t)}
\end{equation}
where $u(\vec{s}_1,..,\vec{s}_N)$ denote the joint probability
distribution of {\em initially} finding the first walker
at a position $\vec{s}_1$, the second at $\vec{s}_2$ and so on.
$\Phi_{1}(\vec{s}_i,t)$ is the survival probability of the
target at time $t$ in the presence of only one walker initially
at position $\vec{s}_{i}$.
In Eq.~(\ref{PhiN(t)}), the sums run over all the lattices sites,
and become in integrals over the space in the case of diffusion
in the continuous space.

From the survival probability, we define the target lifetime density
(TLD), $F_N(t)$, in the presence of $N$ walkers, in the standard way
by
\begin{equation}
F_N(t) = -\frac{d}{dt} \Phi_{N}(t) \,.
\label{FN(t)}
\end{equation}
Then, we can write~\cite{050,060}
\begin{equation}
F_N(t) = \sum_{\vec{s}_1} \ldots \sum_{\vec{s}_N}
u(\vec{s}_1,..,\vec{s}_N)
\,\sum_{i=1}^N F_1(\vec{s}_i,t)
\prod_{j\neq i}^{N}{\Phi_1(\vec{s}_j,t)} ,
\label{FN}
\end{equation}
where $F_1(\vec{s}_i,t)$ is the target lifetime density
in the presence of only one walker, initially at position
$\vec{s}_{i}$. This quantity is defined from
$\Phi_{1}(\vec{s}_i,t)$ in an analogous way to Eq.~(\ref{FN(t)}),
\begin{equation}
F_1(\vec{s}_i,t) = -\frac{d}{dt} \Phi_1(\vec{s}_i,t) \,.
\label{F1}
\end{equation}
In the case of perfect trapping, $F_1(\vec{s}_i,t)$ is the
first-passage time density of the walker.
When $F_1(\vec{s}_i,t)$ is normalized, trapping is certain and
the process is called {\em recurrent} in the sense proposed by
Hughes~\cite{135}. On the other hand, if
\begin{equation}
f_1(\vec{s}_i) = \int_0^{\infty}F_{1}(\vec{s}_i,t)dt < 1 \,,
\label{f1}
\end{equation}
then  the process is called {\em transient}~\cite{135}.
$f_1(\vec{s}_i)$ is the probability that a walker starting
from site $\vec{s}_i$ will ever reach the origin.

Now, using TLD, we introduce the Mean Target Lifetime
(MTL)~\cite{020,140}
\begin{equation}
T_N = \int_0^{\infty} t \,F_{N}(t) dt \,.
\label{MTL1}
\end{equation}
If $t\,\Phi_N(t)\rightarrow 0$ for $t\rightarrow \infty$,
then we can also write
\begin{equation}
T_N = \int_0^{\infty} \Phi_{N}(t) dt \,.
\label{MTL2}
\end{equation}
%

%%%%%%%%%%%%%%%%%%%%%%%%%%%%%%%%%%%%%%%%%%%%%%%%%%%%%%%%%%%%%%%%%%%%%%
\section{First-Passage time}
\label{sec:FPT}
%%%%%%%%%%%%%%%%%%%%%%%%%%%%%%%%%%%%%%%%%%%%%%%%%%%%%%%%%%%%%%%%%%%%%%

A general expression for  $\Phi_{1}(\vec{s}_i,t)$ can be constructed
in terms of the conditional probability $q(\vec{s},t|\vec{s}_i,t=0)$,
corresponding to a walker be in $\vec{s}$ at time $t$, given that
it was at $\vec{s}_i$ at $t=0$, restricted by the presence of
a trap at the origin
\begin{equation}
\Phi_{1}(\vec{s}_i,t)=\sum_{\vec{s}}q(\vec{s},t|\vec{s}_i,t=0) \,,
\label{survi1}
\end{equation}
where the sum runs over all lattice sites and must be replaced
by an integral in the continuous case. This expression is valid
for any kind of trap, allowing for example imperfect trapping
or dynamical gated trapping.

For Markov processes, in the perfect trapping case, we can
additionally exploit the connection between the probability
density of first arrival at the origin at time $t$ from the
initial site $\vec{s}_i$, $F_1(\vec{s}_i,t)$, and the conditional
probability of finding an unrestricted walker at site $\vec{s}$
at time t, given that it was initially at
$\vec{s}_i$, $P(\vec{s},t|\vec{s}_i,t=0)$,~\cite{141}
\begin{equation}
P(\vec{0},t|\vec{s}_i,t=0) =
\Psi(\vec{s}_i,t) \,\delta_{\vec{s}_i,\vec{0}} \;+
\int_{0}^{t} P(\vec{0},t|\vec{0},t') F_1(\vec{s}_i,t') \,dt' \,,
\label{fundamental}
\end{equation}
where $\Psi(\vec{s}_i,\tau)$ is the sojourn probability, i.e.,
the probability that the walker remains on the site $\vec{s}_i$
a time lag $\tau$ without a transition.
Moreover, for an stationary process we have
$P(\vec{s},t|\vec{0},t') = P(\vec{s},t-t'|\vec{0},t=0)$
and the integral in Eq.~(\ref{fundamental}) becomes in a convolution.
Thus, the Laplace transform of Eq.~(\ref{fundamental}) lead us to
the Laplace transform of the first-passage time density
\begin{equation}
\hat{F}_1(\vec{s}_i,u) =
\frac{\hat{P}(\vec{0},u|\vec{s}_i,t=0) -
\hat{\Psi}(\vec{s}_i,u)\delta_{\vec{s}_i,\vec{0}}}
{\hat{P}(\vec{0},u|\vec{0},t=0)} \,,
\label{F1(u)}
\end{equation}
where the caret denotes the Laplace transform of the corresponding
function.
Therefore, using Eqs.~(\ref{F1}) and~(\ref{F1(u)}),
the initial condition $\Phi_{1}(\vec{s}_i,t=0)=1$,
and taking $\vec{s}_i \neq \vec{0}$, we finally get
the Laplace transform of the survival probability of
the target in presence of only one walker,
\begin{equation}
\hat{\Phi}_1(\vec{s}_i,u) =
\frac{\hat{P}(\vec{0},u|\vec{0},t=0)-\hat{P}(\vec{0},u|\vec{s}_i,t=0)}
{u \;\hat{P}(\vec{0},u|\vec{0},t=0)} \,.
\label{Phi(u)}
\end{equation}
Finally, in a similar way as in Eq.(\ref{MTL2}),
the mean first-passage time (MFPT) for the walker can be written as
\begin{equation}
T = T_1 = \int_0^{\infty} \Phi_{1}(t) \,dt \,.
\label{T1}
\end{equation}
%

%%%%%%%%%%%%%%%%%%%%%%%%%%%%%%%%%%%%%%%%%%%%%%%%%%%%%%%%%%%%%%%%%%%%%%
\section{MTL in $d$--dimensions}
\label{sec:d-dim}
%%%%%%%%%%%%%%%%%%%%%%%%%%%%%%%%%%%%%%%%%%%%%%%%%%%%%%%%%%%%%%%%%%%%%%

If the trapping is perfect, MTL is also the MFPT for the first
of the set of $N$ random walkers in reach the target.
An interesting problem is presented when the MFPT for only
one walker diverges, i.e., when the integral of Eq.~(\ref{T1})
diverges.
In this situation, the interesting question that arises is to find
the minimum number of walkers such that MTL becomes finite~\cite{080},
independently of the initial positions of the walkers.
That is, from Eqs.~(\ref{PhiN(t)}) and~(\ref{MTL2}), to find $N$
for which the integral
\begin{equation}
\int_0^{\infty} \prod_{i=1}^{N}\Phi_{1}(\vec{s}_i,t) \,dt \,,
\label{TN}
\end{equation}
converges.
Using our formalism, we can directly rederive the known
results~\cite{080} for this problem.

For concreteness, in this section, we use the CTRW~\cite{150,160}
for the walker's dynamics. This allows us compute analytically
the survival probability of the trap in presence of one walker.
Hence, we can write
\begin{equation}
\hat{P}(\vec{s},u|\vec{0},t=0) = \frac{1-\hat{\psi}(u)}{u} \;
G(\vec{s},\hat{\psi}(u)) \,,
\label{CTRW}
\end{equation}
where $\hat{\psi}(u)$ is the Laplace transform of the pausing time
probability density, $\psi(t)$, and $G(\vec{s},z)$ is the lattice
Green's function. In d--dimensions, it is given by
\begin{equation}
G(\vec{s},z) = \frac{1}{(2 \pi)^d}
\int_{-\pi}^{\pi} \cdots \int_{-\pi}^{\pi}
\frac{\exp(-i \vec{s}\cdot\vec{k})}{1-z \,\Lambda(\vec{k})}
\;d^dk \,,
\label{Green}
\end{equation}
where $\Lambda(\vec{k})$ is the structure function of the lattice.
For a symmetrical walk on a simple cubic d--dimensional lattice we
get
%
%\begin{equation}
$\Lambda(\vec{k}) = (\cos k_1 + \dots + \cos k_d)/d$,
%\label{structure}
%\end{equation}
%
where $k_i$ is the i-th component of $\vec{k}$.

Assuming that $\hat{\psi}(u) \rightarrow 1$ for $u \rightarrow 0$,
the behavior of $G(\vec{s},\hat{\psi}(u))$ is given by the values
of $\vec{k}$ such that $\Lambda(\vec{k}) \approx 1$, i.e.,
$|\vec{k}| << 1$. Thus, $\cos k_i \approx 1 -k_i^2/2$
and $\Lambda(\vec{k}) = 1 - |\vec{k}|^2/(2d)$.
Therefore, we only need to consider the expansion
$\exp(-i \vec{s}\cdot\vec{k}) \approx 1 -i \vec{s}\cdot\vec{k}
- (\vec{s}\cdot\vec{k})^2/2$ in the numerator of Eq.~(\ref{Green}).
Hence, for $u \rightarrow 0$, we get $G(\vec{s},\hat{\psi}(u))
\approx G(\vec{0},\hat{\psi}(u)) - K_d(\vec{s},\hat{\psi}(u))$,
where
\begin{equation}
\begin{array}{ll}
K_d(\vec{s},\hat{\psi}(u)) = &
\displaystyle
\frac{1}{2 \,(2 \pi)^d} \times \\ &
\displaystyle
\int_{-\pi}^{\pi} \cdots \int_{-\pi}^{\pi}
\frac{(\vec{s}\cdot\vec{k})^2 \;d^dk}
{1-\hat{\psi}(u) + \hat{\psi}(u) \,k^2/(2d)}
\,.
\end{array}
\label{Kd}
\end{equation}
Given that we are dealing with unrestricted and spatially homogeneous
walks, we get
$P(\vec{s},t|\vec{s}_i,t=0) = P(\vec{s}-\vec{s_i},t|\vec{0},t=0)$.
Hence, using these results in Eq.~(\ref{Phi(u)}), we obtain
\begin{equation}
\hat{\Phi}_1(\vec{s}_i,u) \approx
\frac{K_d(\vec{s_i},\hat{\psi}(u))}{u \,G(\vec{0},\hat{\psi}(u))} \,,
\label{Phi1}
\end{equation}
where we have used Eq.~(\ref{CTRW}) and the translational invariance
of a CTRW on unbounded lattices
($P(\vec{0},t|\vec{s},t=0) = P(\vec{s},t|\vec{0},t=0)$).

$K_d(\vec{s},z)$ remains finite when $z=1$ since the $k^2$
in the numerator just cancels the singularity in the denominator.
On the other hand, the singularity at $z=1$ in $G(\vec{0},z)$
depends on the dimensionality of the lattice.
For studying this singularity, a good approximate expression of
Eq.~(\ref{Green}) is obtained taking the integrals over the
d--ball of radius $\pi$ inscribed into the first Brillouin
zone. Thus, for $u \rightarrow 0$, we get
\begin{equation}
G(\vec{0},\hat{\psi}(u)) \approx \frac{d \,C_d}{(2 \pi)^d}
\int_{0}^{\pi}
\frac{k^{d-1} \;dk}{1-\hat{\psi}(u) + \hat{\psi}(u) \,k^2/(2d)}
 \,,
\label{Green0}
\end{equation}
where $C_d = \pi^{d/2} / \Gamma(d/2+1)$.

For normal diffusion, the pausing time density is
$\psi(t) = \lambda \exp(-\lambda t)$~\cite{150}. Hence,
\begin{equation}
\hat{\psi}(u) = \left( 1+ \frac{u}{\lambda} \right)^{-1} \,,
\label{psi}
\end{equation}
and we get
$\hat{\psi}(u) \approx 1 - {u}/{\lambda}$, for $u\rightarrow 0$.
For anomalous diffusion, this asymptotic behavior is generalized
by $ \hat{\psi}(u) \approx 1 - \kappa \,u^{\alpha}$,
with  $0 < \alpha < 1$. Hence, normal diffusion is the most
severe case when we analyse the divergence of
$G(\vec{0},z)$ at $z=0$.

%%%%%%%%%%%%%%%%%%%%%%%%%%%%%%%%%%%%%%%%%%%%%%%%%%%%%%%%%%%%%%%%%%%%%%
\subsection{One dimension}
\label{sub:d=1}
%%%%%%%%%%%%%%%%%%%%%%%%%%%%%%%%%%%%%%%%%%%%%%%%%%%%%%%%%%%%%%%%%%%%%%

For $d=1$, $C_1=2$, and by direct integration of Eq.~(\ref{Green0})
we obtain
\begin{equation}
%\begin{array}{ll}
G(\vec{0},\hat{\psi}(u)) \approx
%&
\displaystyle \frac{1}{\pi}
\frac{1}{\sqrt{(1-\hat{\psi}(u))\hat{\psi}(u)/2}}
%\,\times \\
%&
\; \arctan \left(
\pi \displaystyle \sqrt{\frac{\hat{\psi}(u)/2}{1-\hat{\psi}(u)}}
\right) \,.
%\end{array}
\label{Green1}
\end{equation}
For $u \rightarrow 0$,
$\hat{\psi}(u)/(1-\hat{\psi}(u)) \rightarrow \infty$
and the last factor of Eq.~(\ref{Green1}) goes to $\pi/2$.
Thus, for normal diffusion,
\begin{equation}
G(\vec{0},\hat{\psi}(u)) \approx
\frac{1}{\sqrt{2 \,(1-\hat{\psi}(u))}} \propto u^{-1/2} \,.
\label{Green1d}
\end{equation}
In this manner, from Eq.~(\ref{Phi1}), we obtain
$\hat{\Phi}_1(\vec{s}_i,u) \propto u^{-1/2}$ and
using a Tauberian theorem~\cite{140,160}, we immediately get
$\Phi_1(\vec{s}_i,t) \propto t^{-1/2}$.
Therefore, Eq.~(\ref{T1}) diverges and the convergence of
Eq.~(\ref{TN}) is obtained for $N>2$.
This result is also valid for imperfect traps~\cite{060,100},
and for dynamical gated trapping~\cite{110,120} too.

%%%%%%%%%%%%%%%%%%%%%%%%%%%%%%%%%%%%%%%%%%%%%%%%%%%%%%%%%%%%%%%%%%%%%%
\subsection{Two dimensions}
\label{sub:d=2}
%%%%%%%%%%%%%%%%%%%%%%%%%%%%%%%%%%%%%%%%%%%%%%%%%%%%%%%%%%%%%%%%%%%%%%

For $d=2$, $C_2=\pi$, and by direct integration of Eq.~(\ref{Green0})
we obtain
\begin{equation}
G(\vec{0},\hat{\psi}(u)) \approx
\frac{1}{\pi \,\hat{\psi}(u)}
\ln \left(
1 + \frac{\pi^2 \,\hat{\psi}(u)}{4 (1-\hat{\psi}(u))}
\right) \,.
\label{Green2}
\end{equation}
Thus, for $u \rightarrow 0$,
$G(\vec{0},\hat{\psi}(u)) \approx -\ln(1-\hat{\psi}(u))/\pi
\propto -\ln u$, $\hat{\Phi}_1(\vec{s}_i,u) \propto - 1/ (u \,\ln u)$
and $\Phi_1(\vec{s}_i,t) \propto \ln t$.
Therefore, the convergence of Eq.~(\ref{TN}) is not reached
for any value of $N$.

%%%%%%%%%%%%%%%%%%%%%%%%%%%%%%%%%%%%%%%%%%%%%%%%%%%%%%%%%%%%%%%%%%%%%%
\subsection{$\bf d \geq 3$}
\label{sub:d>2}
%%%%%%%%%%%%%%%%%%%%%%%%%%%%%%%%%%%%%%%%%%%%%%%%%%%%%%%%%%%%%%%%%%%%%%

For $d>2$, $G(\vec{0},\hat{\psi}(u))$ remains finite
for $u \rightarrow 0$ since the $k^{d-1}$ in the numerator just
cancels the singularity in the denominator of Eq.~(\ref{Green0}).
Thus, from Eqs.~(\ref{Phi1}),
$\hat{\Phi}_1(\vec{s}_i,u) \propto 1/u$.
Alternatively this behavior can be seen from Eq.~(\ref{F1}),
and taking into account Eq.~(\ref{f1}).
Thus, the asymptotic behavior of the survival probability results
\begin{eqnarray}
\Phi_1(\vec{s}_i,t\rightarrow\infty) =
1 - \int_0^{t\rightarrow\infty} F_1(s_i,t') dt'
=  1 - f_1(\vec{s}_i)\,.
\end{eqnarray}
Therefore, the long time behavior of the survival probability
is constant, i.e., time independent.
For $d\geq3$, the process results transient~\cite{135}
($0 < f_1(\vec{s}_i) < 1$) so each factor in Eq.~(\ref{TN})
results $0 < 1 - f_1(s_i) < 1$, and then, the convergence
of Eq.~(\ref{TN}) is not reached for any value of $N$.

On the other hand, although the MFPT diverges for $d\geq3$,
the asymptotic limit of the survival probability  of the target,
which is proportional to $\prod_{i=1}^N(1-f_1(\vec{s}_i))$ with each
factor less than one, decreases monotonously with $N$, independently
of the initial distribution of the walkers.

Alternatively, we can proceed as in Ref.~\cite{080} choosing only
the random walkers that will ever reach the target. For this purpose,
it is necessary to define the conditional probability density of
first arrival at the origin at time $t$ from the initial site
$\vec{s}_i$, given that the walker will eventually arrive there:
$F_1(\vec{s}_i,t)/f_1(\vec{s}_i)$. Working with this quantity,
we can find in $d=3$ an expression equivalent to Eq.~(\ref{Green1d}).
Thus, in $d=3$, MTL is finite if at least three walkers eventually
reach the origin (target).

%%%%%%%%%%%%%%%%%%%%%%%%%%%%%%%%%%%%%%%%%%%%%%%%%%%%%%%%%%%%%%%%%%%%%%
\section{Initial joint distribution}
\label{sec:initial}
%%%%%%%%%%%%%%%%%%%%%%%%%%%%%%%%%%%%%%%%%%%%%%%%%%%%%%%%%%%%%%%%%%%%%%

Now, we consider the effects on TLD and MTL due to different initial
probability distributions, $u(\vec{s}_1,..,\vec{s}_N)$. Most of cases
in the literature~\cite{170,180,190} belong to the following kinds of
distributions:

%%%%%%%%%%%%%%%%%%%%%%%%%%%%%%%%%%%%%%%%%%%%%%%%%%%%%%%%%%%%%%%%%%%%%%
\subsection{Concentrated}
\label{sub:CON}
%%%%%%%%%%%%%%%%%%%%%%%%%%%%%%%%%%%%%%%%%%%%%%%%%%%%%%%%%%%%%%%%%%%%%%

All walkers can begin at the same point of the space,
\begin{equation}
u(\vec{s}_1,..,\vec{s}_N) = \prod_{i=1}^N\delta_{\vec{s}_i,\vec{s}_0}
\,,\;\; (\vec{s}_0 \neq \vec{0})
\,.
\label{con}
\end{equation}
Thus, from Eq.~(\ref{PhiN(t)}) we obtain
\begin{equation}
\Phi_N^{con}(t)= \left( \Phi_1(\vec{s}_0,t) \right)^N \,,
\label{Phi_con}
\end{equation}
and using Eq.~(\ref{FN}) results
\begin{equation}
F_N^{con}(t) = N \, F_1(\vec{s}_0,t) \,
\left( \Phi_1(\vec{s}_0,t) \right)^{N-1} \,.
\end{equation}
Recalling that
$\Phi_1(\vec{s}_0,t) = 1-\int_0^t F_1(\vec{s}_0,\tau) d\tau$,
the last equation may be compared with Eq.~(19) in Ref.\cite{190}.

%%%%%%%%%%%%%%%%%%%%%%%%%%%%%%%%%%%%%%%%%%%%%%%%%%%%%%%%%%%%%%%%%%%%%%
\subsection{Equally likely sites}
\label{sub:ELS}
%%%%%%%%%%%%%%%%%%%%%%%%%%%%%%%%%%%%%%%%%%%%%%%%%%%%%%%%%%%%%%%%%%%%%%

Alternatively, the initial site of each walker can be chosen
by chance among the $M$ sites of a given set $\cal S$ with
equal probability,
\begin{equation}
u(\vec{s}_1,..,\vec{s}_N) = \left\{
\begin{array}{ll}
M^{-N} & \mbox{ if all } \vec{s}_i \in {\cal S}, \\
0      & \mbox{ otherwise,}
\end{array}
\right.
\label{els}
\end{equation}
where ${\cal S}$ is such that $\vec{0} \notin {\cal S}$.
Thus, from Eq.~(\ref{PhiN(t)}) results
\begin{equation}
\Phi_N^{els}(t) =
\left( \frac{1}{M} \sum_{\displaystyle \vec{s} \in {\cal S}}
\Phi_1(\vec{s},t) \right)^N \,.
\label{Phi_els}
\end{equation}
Also, we can write Eq.~(\ref{Phi_els}) as
$\Phi_N^{els}(t) = \left( \left< \Phi_1(\vec{s},t) \right> \right)^N$,
where $\left< \cdots \right>$ denotes the spacial average taken
in the set $\cal S$.
Alternatively, we can recast Eq.~(\ref{Phi_els}) as
\begin{equation}
\Phi_N^{els}(t) =
\left( 1-\frac{1}{M}\sum_{i=1}^M (1-\Phi_1(\vec{s}_i,t)) \right)^N,
\label{Phi_els2}
\end{equation}
where $\vec{s}_i$ ($i=1,\ldots,M$) are the positions of the
$M$ sites of set $\cal S$.
Note that there is not any restriction between $N$ and $M$.
Eq.~(\ref{Phi_els2}) allows us to take the limits
$N \rightarrow \infty$, $M \rightarrow \infty$,
with $N/M \rightarrow \beta$ constant, i.e., the {\em bulk limit}.
In this case we get
%
%\begin{equation}
%\Phi_{\beta}^{els}(t) = \exp \left( -\beta \sum_{\vec{s}\neq\vec{0}}
%\left( 1 - \Phi_1(\vec{s}_i,t) \right) \right) \,,
%\label{Phi_beta}
%\end{equation}
%
\begin{equation}
\Phi_{\beta}^{els}(t) = \exp \left( -\beta S(t) \right) \,,
\label{Phi_beta}
\end{equation}
where
\begin{equation}
S(t)= \sum_{\vec{s}\neq\vec{0}} ( 1 - \Phi_1(\vec{s}_i,t)) \,,
\label{S(t)}
\end{equation}
is the average number of lattice points visited,
at least once time, by one walker, until time $t$.
We are assuming that the series in Eq.~(\ref{S(t)}) converges.

In the continuous d--dimensional space, assuming that $\cal S$
has a finite volume V and that initially each walker begin
uniformly distributed in $V$, the generalization of
Eq.~(\ref{Phi_els}) is immediate,
\begin{equation}
\Phi_N^{els}(t) =
\left( \frac{1}{V} \int_{\cal S} \Phi_1(\vec{s},t) \,d^ds \right)^N\,.
\end{equation}
If the number of walkers per unit of volume, $c=N/V$, is constant,
then in the limit $N \rightarrow \infty$ and  $V \rightarrow \infty$
we also obtain
\begin{equation}
\Phi_c^{els}(t) = \exp \left( -c
\int (1 - \Phi_1(\vec{s},t)) \,d^ds \right) \,,
\label{Phi_c}
\end{equation}
where we assume that the integral over the whole space in bounded.

%%%%%%%%%%%%%%%%%%%%%%%%%%%%%%%%%%%%%%%%%%%%%%%%%%%%%%%%%%%%%%%%%%%%%%
\section{Illustrations}
\label{sec:illustrations}
%%%%%%%%%%%%%%%%%%%%%%%%%%%%%%%%%%%%%%%%%%%%%%%%%%%%%%%%%%%%%%%%%%%%%%

%%%%%%%%%%%%%%%%%%%%%%%%%%%%%%%%%%%%%%%%%%%%%%%%%%%%%%%%%%%%%%%%%%%%%%
\subsection{Finite and semi--infinite chain}
\label{sub:FySI}
%%%%%%%%%%%%%%%%%%%%%%%%%%%%%%%%%%%%%%%%%%%%%%%%%%%%%%%%%%%%%%%%%%%%%%

\begin{figure}[tbp]
\begin{center}
\includegraphics[clip,width=0.90\textwidth]{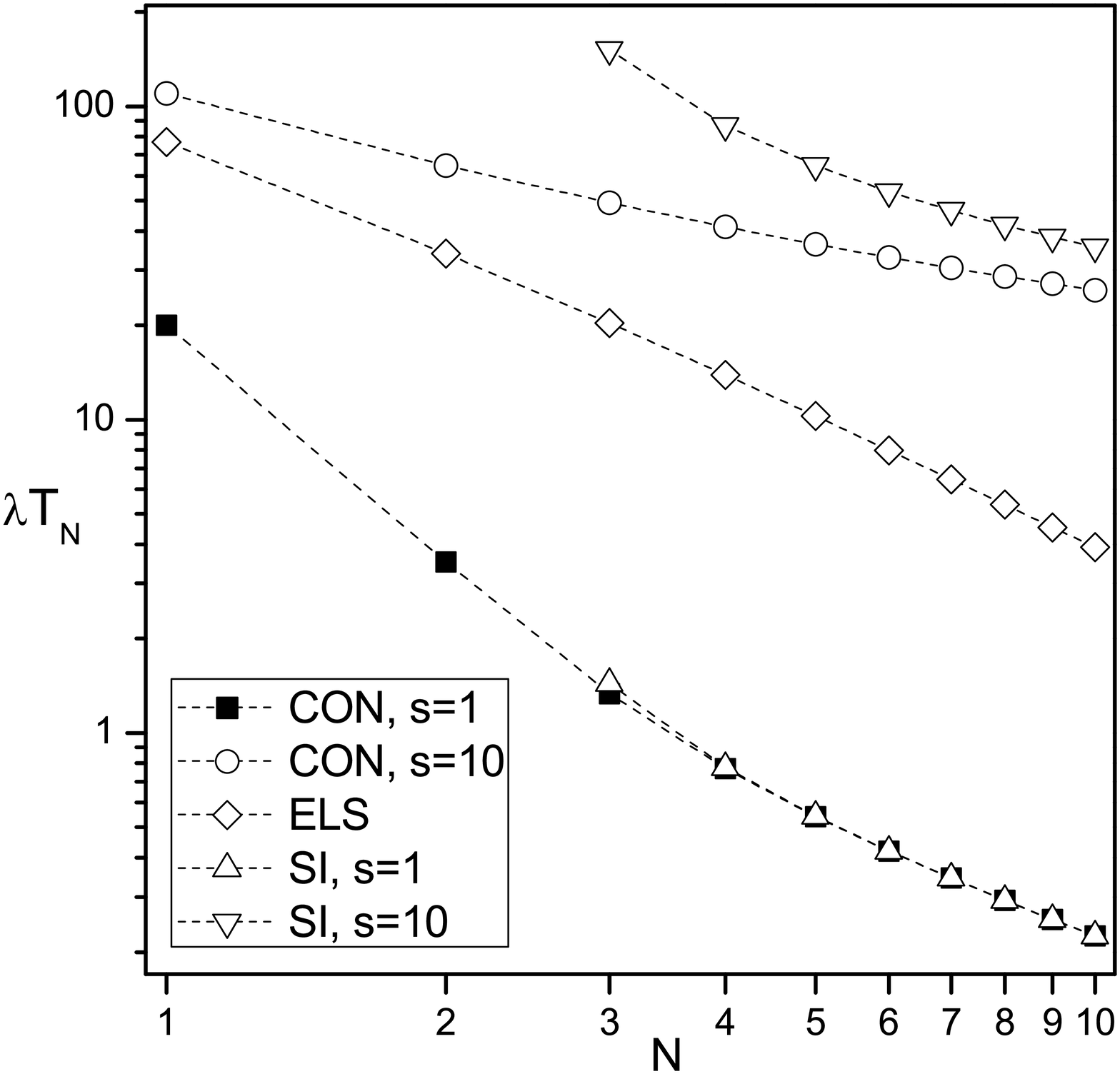}
\caption{$T_N$ as a function of the number $N$ of walkers for
a chain with the trap at the origin. The finite case corresponds
to $L=10$ sites and two different initial distributions:
Concentrated (CON), at site $s=1$ or site s=10; and with all
sites equally likely (ELS). The semi--infinite case (SI) corresponds
to all walkers initially concentrated, at site $s=1$ or site $s=10$.
The dotted lines are only to guide the eye.}
\label{Fig.1}
\end{center}
\end{figure}

As illustration, we now compute the MTL for a perfect trap at the
origin of a chain, in presence of $N$ walkers that jump from any
site to its nearest neighbor with transition rate $\lambda$.
For a finite chain of L sites with absorbing end at the origin
and reflecting end at site L, there is an exact expression
for the Laplace transform of the first-passage time density~\cite{200}
\begin{equation}
\hat{F}_1(j,u) =
\frac{R(u)^j + R(u)^{2L+1-j}}{1 + R(u)^{2L+1}} \,,
\end{equation}
($j=1,2,\ldots, L$) where
$R(u) = \left( r + 1  - \sqrt{r^2 + 2 \,r} \right)$
and $r=u/\lambda$.
This expression can be Laplace antitransformed
in exact way~\cite{210} but it is rather clumsy to display here.
Moreover, for use Eqs.~(\ref{Phi_con}) or~(\ref{Phi_els}), we need
previously make the integration involved in Eq.~(\ref{FN(t)})
and later make the integration in Eq.~(\ref{MTL2}). We performed
numerically these integrals~\cite{210}.
In Fig.~\ref{Fig.1} we plot the values of $T_N$ for a chain
with $L=10$ and initial distribution of walkers given by
Eqs.~(\ref{con}) and~(\ref{els}).
Strikingly, we find in both cases a power-law behaviour
(see quasi--linear relation in the log-log plot) for almost
all values of $N$.

\begin{figure}[tbp]
\begin{center}
\includegraphics[clip,width=0.90\textwidth]{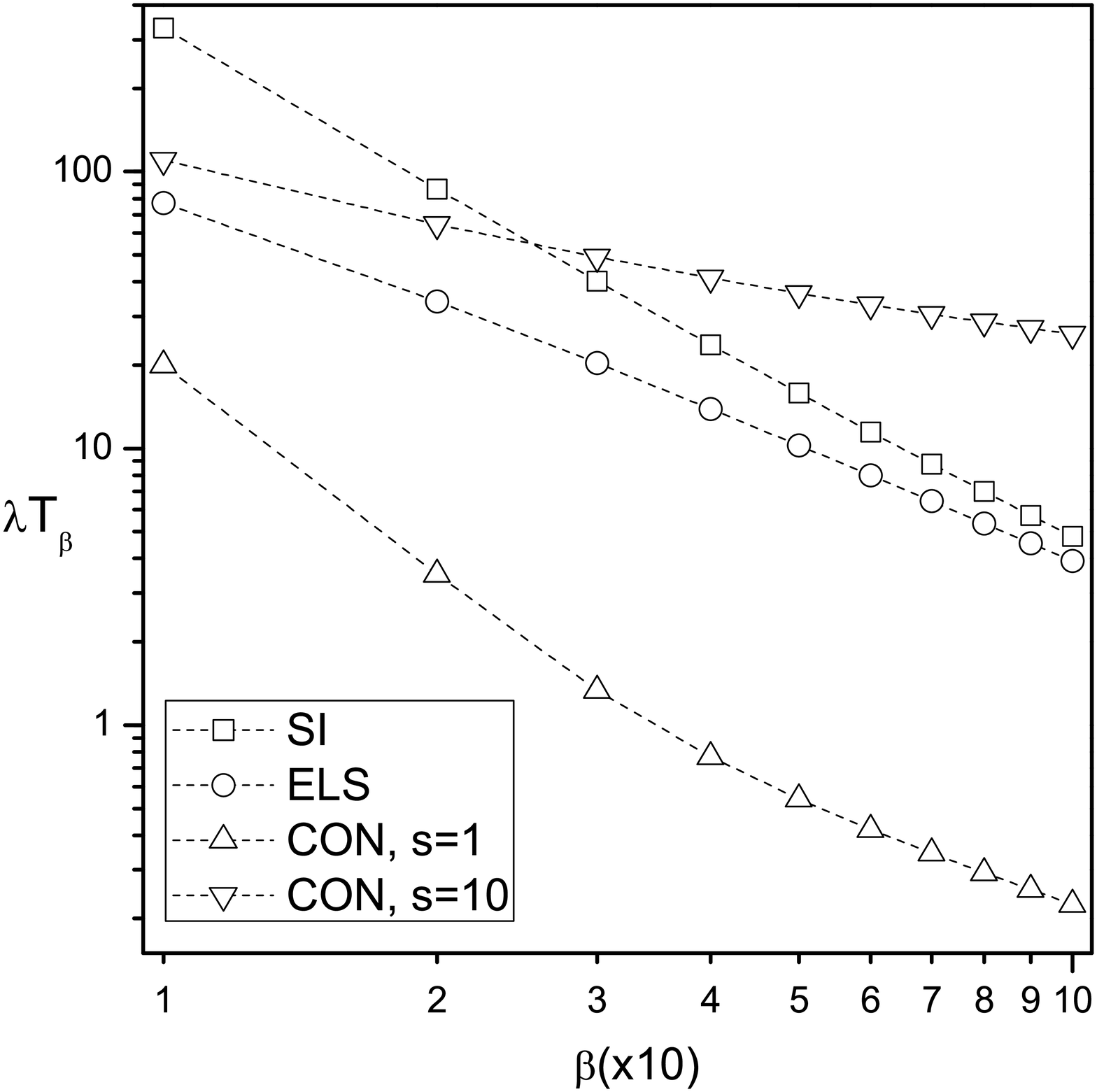}
\caption{$T_{\beta}$ as a function of the concentration of
walkers $\beta$, for a chain with the target at the origin.
The semi--infinite case (SI) corresponds to the bulk limit.
The finite case corresponds to $L=10$ sites ($\beta=N/L$)
and all walkers initially concentrated (CON), at site $s=1$
or site $s=10$; or with all sites equally likely (ELS).
The dotted lines are only to guide the eye.}
\label{Fig.2}
\end{center}
\end{figure}

If the chain is semi--infinite, an explicit and simple expression
for $\Phi_1(j,t)$ ($j=1,2,\ldots$) can be derive from results in
the literature~\cite{220,060}
\begin{equation}
\Phi_1(j,t) = e^{-\lambda t} \,(I_0(\lambda t)-I_j(\lambda t))
+ 2 \,e^{-\lambda t} \,\sum_{k=1}^{j} I_k(\lambda t) \,,
\end{equation}
where $I_k(x)$ are the modified Bessel functions.
Using this expression in  Eqs.~(\ref{Phi_con}) or~(\ref{Phi_beta})
and integrating numerically~\cite{210} in Eq.~(\ref{MTL2}),
we can evaluate the MTL for the initial distributions of
Eqs.~(\ref{con}) and~(\ref{els}). Figure~\ref{Fig.2}
plots the situation for walkers initially distributed
with concentration $\beta$ on the chain.

For comparison purposes, we also include in Fig.~\ref{Fig.1}
the plots corresponding to a semi--infinite chain with
all walkers initially concentrated, at site $s=1$ or site $s=10$.
In the later plots, the minimum number of walkers to obtain finite
values of $T_N$ is three, as has been quoted in Sec.~\ref{sub:d=1}.
We want to stress the notorious resemblance obtained, for not
so large values of $N$, between the cases of finite chain and
semi--infinite chain for all walkers initially concentrated.
Fig.~\ref{Fig.2} also includes the plots corresponding to a finite
chain of $L=10$ sites. In this graph the noticeable resemblance
for large values of $\beta$ is given between the cases of
semi--infinite chain and finite chain with initial distribution
of equally likely sites.

%%%%%%%%%%%%%%%%%%%%%%%%%%%%%%%%%%%%%%%%%%%%%%%%%%%%%%%%%%%%%%%%%%%%%%
\subsection{Bulk limit in d--dimensions}
\label{sub:Bulk}
%%%%%%%%%%%%%%%%%%%%%%%%%%%%%%%%%%%%%%%%%%%%%%%%%%%%%%%%%%%%%%%%%%%%%%

For d--dimensional lattices with a perfect trap at the origin,
the bulk limit is given by Eq.~(\ref{Phi_beta}).
Using Eq.~(\ref{MTL2}), MTL can be written in this case as
\begin{equation}
T_{\beta} = \int_{0}^{\infty}
\,\exp \left( -\beta S(t) \right) \,dt \,.
\label{T_beta}
\end{equation}
It can be shown that the laplace transform of $S(t)$,
asumming a CTRW dynamics, is given by~\cite{230}
\begin{equation}
\hat{S}(u)=\frac{\hat{\psi}(u)}{u^2 \,P(\vec{0},u|\vec{0},t=0)} \,,
\label{S(u)}
\end{equation}
where $\hat{\psi}(u)$ is given by Eq.~(\ref{psi}).

\begin{figure}[tbp]
\begin{center}
\includegraphics[clip,width=0.90\textwidth]{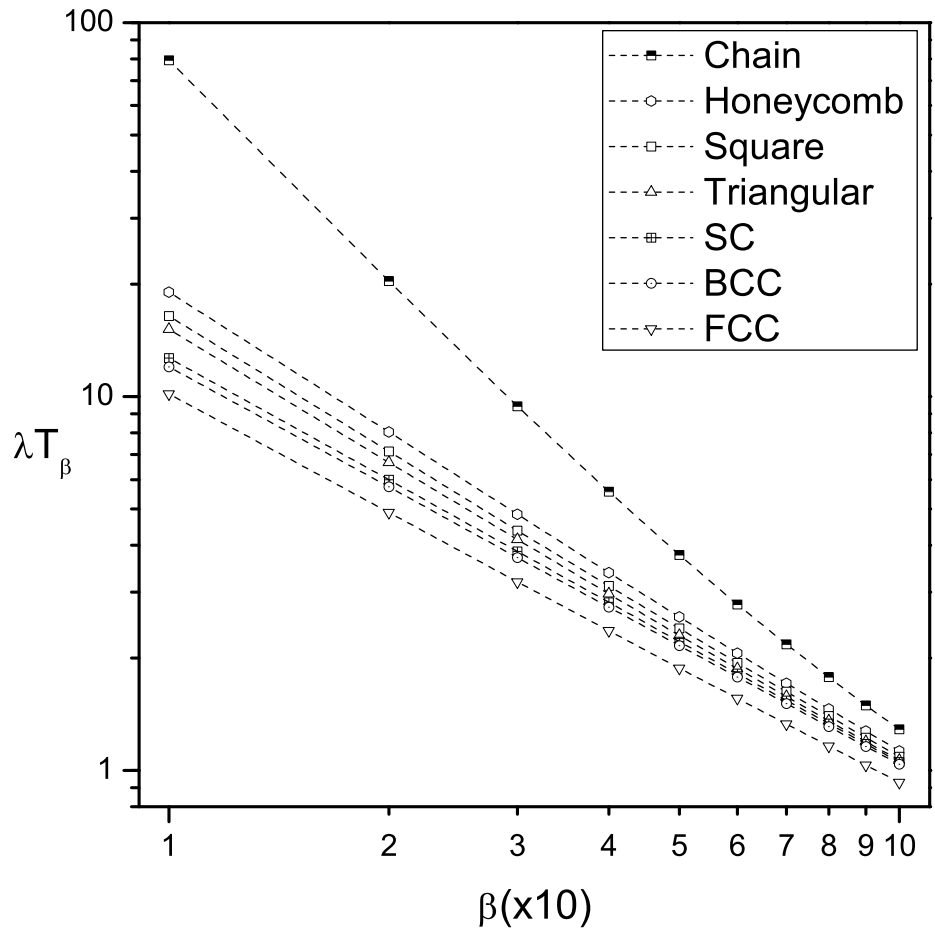}
\caption{$T_\beta$ as a function of the concentration
of walkers, $\beta$, in the bulk limit, for different lattices.
The coordination numbers of the lattices, from top to botton,
are $\kappa = 2,3,4,5,6,8,12$.}
\label{Fig.3}
\end{center}
\end{figure}

Using known expressions for $P(\vec{0},u|\vec{0},t=0)$
given in Ref.~\cite{135}, we plot in Fig.~(\ref{Fig.3})
$T_\beta$ as function of $\beta$ for different lattices
in $d = 1$ (chain), $d = 2$ (honeycomb, square, and triangular),
and $d = 3$ (SC, BCC, and FCC).
$T_\beta$ shows again a quasi--linear relation in the log--log
plot and, as expected, $T_\beta$ monotonously decrease as the
coordination number of the lattice, $\kappa$, is increased.

%%%%%%%%%%%%%%%%%%%%%%%%%%%%%%%%%%%%%%%%%%%%%%%%%%%%%%%%%%%%%%%%%%%%%%
\subsection{Imperfect trapping in the continuous space}
\label{sub:Imperfect}
%%%%%%%%%%%%%%%%%%%%%%%%%%%%%%%%%%%%%%%%%%%%%%%%%%%%%%%%%%%%%%%%%%%%%%

The semi--straight line with an imperfect trap at the origin
is another interesting illustration of our concepts.
The probability density (in presence of the trap), $q(x, t |x_0,t=0)$,
for finding a particle at the location $x$ at time $t$, given that
it departed from site $x_0$ at the time $t=0$, satisfies the classical
diffusion equation
\begin{equation}
\frac{\partial q}{\partial t} = D \,\frac{\partial^2q}{\partial x^2}
\,,
\label{continuo1}
\end{equation}
where D is the diffusion coefficient.
An imperfect trap is described by the radiation boundary
condition~\cite{240}
\begin{equation}
\left. \frac{\partial q}{\partial x} \right|_{x=0} =
\gamma \,q(x=0,t|x_0,t=0) \,,
\label{boundcond}
\end{equation}
where $\gamma$ measure the efficiency of the trap. $\gamma=0$
corresponds to a reflecting boundary and perfect trapping
is reached in the limit $\gamma\rightarrow\infty$.
The solution of Eqs.~(\ref{continuo1}) and~(\ref{boundcond})
can be read from Ref.~\cite{250}
\begin{equation}
q(x, t |x_0,t=0) = q_a - \gamma \,q_b \,,
\end{equation}
where
\begin{equation}
q_a = \frac
{\exp \left( -\displaystyle\frac{(x-x_0)^2}{2Dt} \right)
+
\exp \left( -\displaystyle\frac{(x+x_0)^2}{2Dt} \right)}
{2\sqrt{\pi Dt}}
\end{equation}
and
\begin{equation}
q_b = \exp \left( D\gamma^2t+\gamma(x+x_0) \right)
\mbox{erfc}\left( \frac{x+x_0}{2\sqrt{Dt}}+\gamma
\sqrt{Dt}\right) \,.
\end{equation}

The relation between the survival probability $\Phi_1(x_0,t)$
and $q(x,t|x_0,t=0)$ is given by Eq.~(\ref{survi1}).
Therefore, for the imperfect trapping we get~\cite{260}
\begin{equation}
\Phi_1(x_0,t) = 1 -
\mbox{erfc} \left( \frac{x_0}{2\sqrt{D t}} \right)
+ \exp \left( D \gamma^2 t + \gamma x_0 \right) \,
\mbox{erfc} \left( \frac{x_0}{2\sqrt{Dt}} + \gamma\sqrt{Dt} \right)
\label{Phi_1(x_0,t)}
\,.
\end{equation}
Notice that the expected expression for perfect trapping
($\gamma\rightarrow\infty$) is directly recovered
\begin{equation}
\Phi_1(x_0,t) = 1 - \mbox{erfc} \left( \frac{x_0}{2\sqrt{Dt}} \right)
\,.
\end{equation}

\begin{figure}[tbp]
\begin{center}
\includegraphics[clip,width=0.90\textwidth]{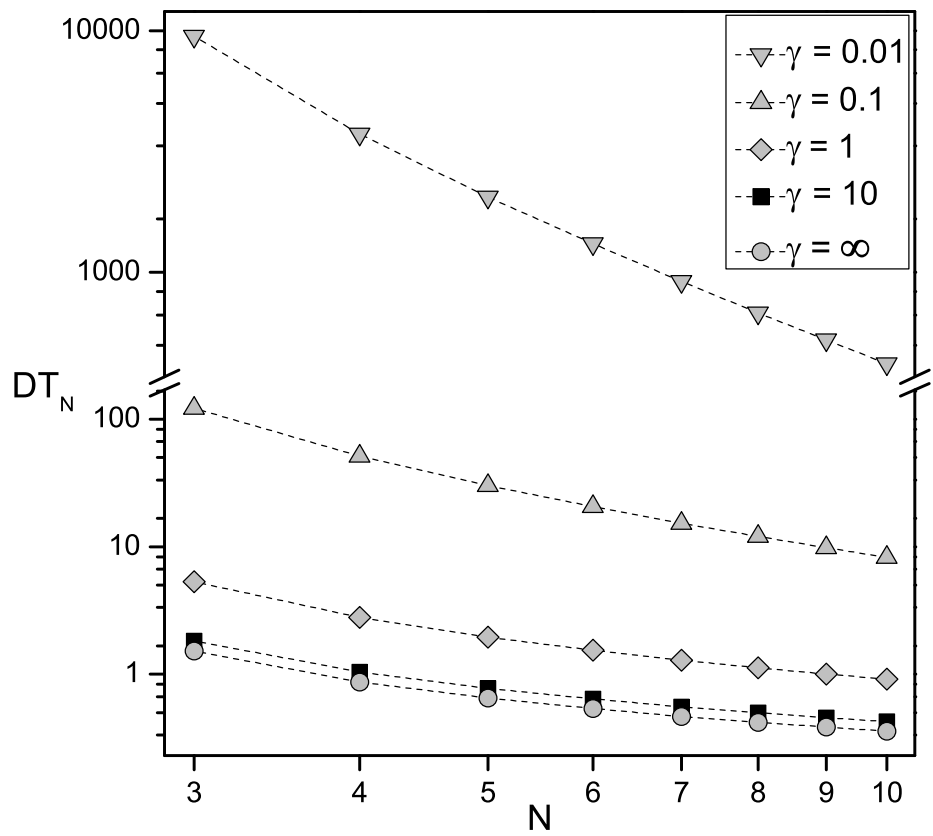}
\caption{$T_N$ as a function of the number $N$ of walkers for
semi--straight line with an imperfect trap at the origin.
The dotted lines are only to guide the eye.}
\label{Fig.4}
\end{center}
\end{figure}

Figure~\ref{Fig.4} graphs $T_N$ for several situations of trapping
efficiency ($\gamma$) and for the initial distribution of walkers
concentrated at $x_0 = 1$.
To make each plot we have used  Eq.~(\ref{Phi_1(x_0,t)})
and~(\ref{Phi_con}), and we have make numerically~\cite{210}
the integration of Eq.~(\ref{MTL2}).
Also in the continuous space, we obtain quasi--linear relations
in the log--log plots of $T_N$ vs $N$ for not so large values of $N$.

%%%%%%%%%%%%%%%%%%%%%%%%%%%%%%%%%%%%%%%%%%%%%%%%%%%%%%%%%%%%%%%%%%%%%%
\section{Target search with intermittent motion}
\label{sec:Intermittent}
%%%%%%%%%%%%%%%%%%%%%%%%%%%%%%%%%%%%%%%%%%%%%%%%%%%%%%%%%%%%%%%%%%%%%%

Examples of intermittent processes may be found in many fields.
For instance, the case of a reactant that freely diffuses in
a solvent and intermittently binds to a cylinder~\cite{122}.
Also, we found intermittent motion in  the binding of a protein
to specific sites on DNA for regulating transcription,
as it is the case when the protein has the ability of diffuse
in one dimension by sliding along the length of the DNA,
in addition to their diffusion in bulk solution~\cite{124}.
Moreover, intermitency could be associate with dynamical
trapping problems~\cite{270}.
We also found the search strategies like those implemented
by animals in the pursuit of prey, or even in human activities
such as victim localization, among the most representative
examples of intermittent motion at macroscopic scales
(see Ref.~\cite{126} and references therein).
Recent works on intermittent search strategies were focused on the
analysis of trapping of a single walker wandering in the presence
of distributed traps. The magnitude that is usually estimated
is this case is the named search time~\cite{128}.

An interesting application of our present formalism
is the related problem of a single static target among
a set of initially uniformly distributed searchers,
switching intermittently between two states of motion~\cite{126,127}.
In what follows, we calculate MTL for a target (perfect trap) at the
origin of an infinite chain. We assume that each walker can be in
either of two propagation states. In one of them, the displacement
of the searchers is a random walk with symmetric jumps to nearest
neighbors sites, with constant rate $\lambda$. In the other state,
the searchers also perform a symmetrical random walk, but jumping
to next-nearest neighbors, with the same constant rate $\lambda$.
Thus, in the second state the diffusion is twice as big
as in the other state, but in the second internal state, the
walker can skip over the target without trapping.
Transitions between the first and the second internal state,
take place with rate constant $\gamma _1$, whereas the opposite
transitions, between the second and first state are at rate
$\gamma _2$.
The coupled master equations that describe this composite process
of one walker are then
\begin{equation}
\displaystyle\frac{\partial P_{1}(j,t)}{\partial t} =
\displaystyle\frac{\lambda}{2} \,
\left( P_{1}(j+1,t) + P_{1}(j-1,t) \right)
- \lambda \,P_{1}(j,t) + \gamma_{2} \,P_{2}(j,t)
- \gamma_{1} P_{1}(j,t) \;,
\end{equation}
\begin{equation}
\displaystyle\frac{\partial P_{2}(j,t)}{\partial t} =
\displaystyle\frac{\lambda}{2} \,
\left( P_{2}(j+2,t) + P_{2}(j-2,t) \right)
- \lambda \,P_{2}(j,t) + \gamma_{1} \,P_{1}(j,t)
- \gamma_{2} P_{2}(j,t) \;,
\end{equation}
where $P_1(j,t)$~$(P_2(j,t))$ is the joint probability that
the walker be at site $j$ with internal state $1$~$(2)$ at time $t$.
If we assume equally likely sites as initial distribution of walkers
and the bulk limit defined in subsection~\ref{sub:ELS},
we can use Eqs.(\ref{T_beta}) and~(\ref{S(u)}) with
$P(j=0,u)\equiv P_1(j=0,u)+P_2(j=0,u)$.
In this manner, we can compute $T_{\beta}$ as a function
of $\gamma_1$ and $\gamma_2$.

\begin{figure}[tbp]
\begin{center}
\includegraphics[clip,width=0.90\textwidth]{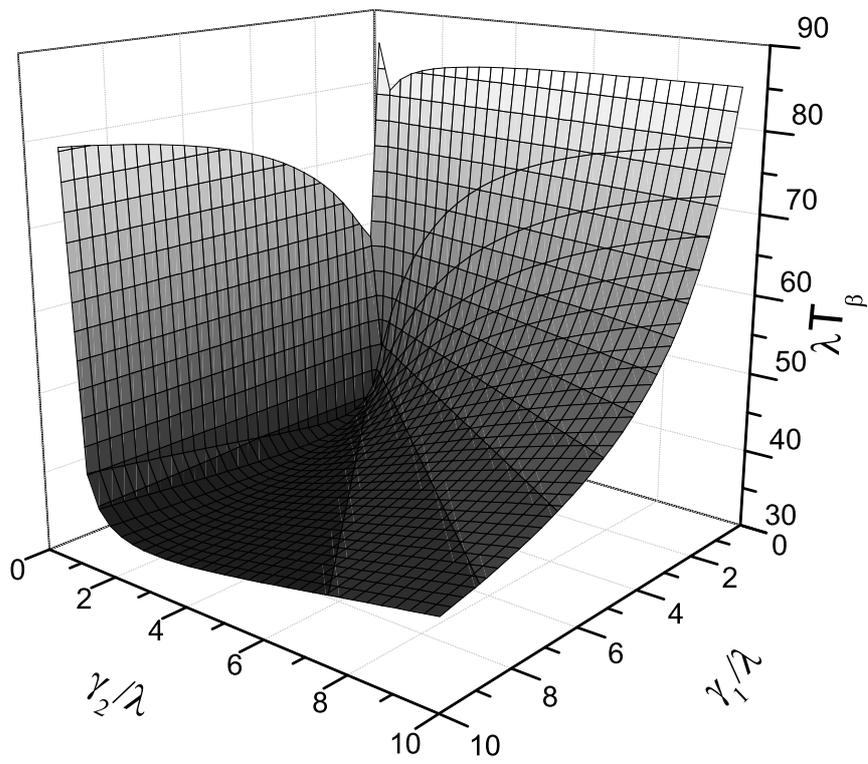}
\caption{$T_{\beta}$ in the bulk limit, for a concentration
$\beta=0.1$ of walkers, as a function
of parameters $\gamma_1$ and $\gamma_2$.}
\label{Fig.5}
\end{center}
\end{figure}

In Fig.~\ref{Fig.5} we draw the behavior of MTL, in the bulk limit,
as a function of the parameters of transition $\gamma_1$
and $\gamma_2$. As can be seen from the figure, we obtain a region
of optimal values in the parameter space $(\gamma_1,\gamma_2)$ which
can be appreciated by the grey scale (darker means a smaller value
in $T_{\beta}(\gamma_1,\gamma_2)$). The valley in the surface
indicates that we can tune the parameters to optimize the search.
Notice how the MTL adequately characterize the improvement
provided by intermittent search strategy~\cite{127}.
Similar behavior has been exhibited in Ref.~\cite{128}
despite the difference in the addressed problems.

%%%%%%%%%%%%%%%%%%%%%%%%%%%%%%%%%%%%%%%%%%%%%%%%%%%%%%%%%%%%%%%%%%%%%%
\section{Conclusions}
\label{sec:fin}
%%%%%%%%%%%%%%%%%%%%%%%%%%%%%%%%%%%%%%%%%%%%%%%%%%%%%%%%%%%%%%%%%%%%%%

This paper provides a simple, general, and unified formalism for
the lifetime statistics of a fixed target in presence of a set
of independent hunters or gatherers that diffuse in the space.
Our framework compresses normal and anomalous diffusion on lattices
as well as in the continuous space. Also, our scheme allows us
to consider perfect an imperfect trapping and can be directly
extended for dynamical traps.
Our main quantity, the MTL was introduced and its connections
with any other physical quantities, relevant for the temporal
statistics of trapping, were established. Particularly,
the role of the initial spatial distribution was discussed.

For trapping problems, where we deal with only one tramp,
MFPT approach is limited when the process is transient
(see Eq.~(\ref{f1})) or the recurrence time in the
lattice is infinite, because the MFPT diverge.
Although MTL overcome this problem in one dimension,
if the number of walkers is increased, it also diverges
for $d>1$.
However, the bulk limit of MTL is finite in all situations.
This robust property points out that MTL is the relevant
physical quantity for situations where we consider the
bulk density of diffusing particles and only one isolated
capture center.

Additionally, a striking feature of MTL is it presents non-universal
scaling laws when it is plotted as a function of the number
or density of walkers.
Moreover, MTL is an efficient global optimizer for search strategies
using intermittent motion. The MTL surface has a valley, allowing
us to tune up the parameters that regulate the intermittency,
and thus to minimize the time of search.

%%%%%%%%%%%%%%%%%%%%%%%%%%%%%%%%%%%%%%%%%%%%%%%%%%%%%%%%%%%%%%%%%%%%%%
%%%%%%%%%%%%%%%%%%          Acknowledgments         %%%%%%%%%%%%%%%%%%
%%%%%%%%%%%%%%%%%%%%%%%%%%%%%%%%%%%%%%%%%%%%%%%%%%%%%%%%%%%%%%%%%%%%%%

\vspace{0.7cm}
{\bf Acknowledgements}
This work was partially supported by grant from
``Se\-cre\-ta\-r{\'\i}a de Cien\-cia y Tec\-no\-lo\-g{\'\i}a
de la Uni\-ver\-si\-dad Na\-cio\-nal de C\'or\-doba''
(Code: 05/B380, Res.\ SeCyT 69/08).

%%%%%%%%%%%%%%%%%%%%%%%%%%%%%%%%%%%%%%%%%%%%%%%%%%%%%%%%%%%%%%%%%%%%%%
%%%%%%%%%%%%%%%%%%             References           %%%%%%%%%%%%%%%%%%
%%%%%%%%%%%%%%%%%%%%%%%%%%%%%%%%%%%%%%%%%%%%%%%%%%%%%%%%%%%%%%%%%%%%%%

\vspace{0.7cm}
{\bf References}
%

%

%%%%%%%%%%%%%%%%%%%%%%%%%%%%%%%%%%%%%%%%%%%%%%%%%%%%%%%%%%%%%%%%%%%%%%
%%%%%%%%%%%%%%%%%%        Figure and Captions       %%%%%%%%%%%%%%%%%%
%%%%%%%%%%%%%%%%%%%%%%%%%%%%%%%%%%%%%%%%%%%%%%%%%%%%%%%%%%%%%%%%%%%%%%

\newpage
\centerline{\bf \LARGE Figure and Captions}
\newpage

%

%%%%%%%%%%%%%%%%%%%%%%%%%%%%%%%%%%%%%%%%%%%%%%%%%%%%%%%%%%%%%%%%%%%%%%
\end{document}